\documentstyle[12pt,epsfig]{article}

\newlength{\dinwidth}
\newlength{\dinmargin}
\def\lapproxeq{\lower .7ex\hbox{$\;\stackrel{\textstyle                                         
<}{\sim}\;$}}                                         
\def\gapproxeq{\lower .7ex\hbox{$\;\stackrel{\textstyle                                         
>}{\sim}\;$}}                                         
       
\newcommand{\be}{\begin{equation}}       
\newcommand{\ee}{\end{equation}}       
\newcommand{\bea}{\begin{eqnarray}}       
\newcommand{\eea}{\end{eqnarray}}

\begin{document}
\begin{flushright}
DTP/00/50 \\
4 August 2000 \\
\end{flushright}

\begin{center}
\vspace*{2cm}
{\Large \bf A new determination of the QED coupling $\alpha (M_Z^2)$} \\

\vspace*{0.5cm}                
{\Large \bf lets the Higgs off the hook}                                                         

\vspace*{1cm}
A.D. Martin$^a$, J. Outhwaite$^a$ and M.G. Ryskin$^{a,b}$ \\

\vspace*{0.5cm}

$^a$ Department of Physics, University of Durham, Durham, DH1 3LE, UK. \\

$^b$ Petersburg Nuclear Physics Institute, 188350, Gatchina,
St.\ Petersburg, Russia.

\end{center}

\vspace*{0.3cm}

\begin{abstract}
We use the available data on $e^+ e^- \rightarrow$ hadrons to determine the value of the 
QED coupling at the $Z$ pole.  The inclusion of recent preliminary BES-II data in the analysis 
is found to reduce the ambiguity in the determination of the coupling, 
particularly that arising from the use 
of either exclusive or inclusive data in the energy region 
$\sqrt{s} \lapproxeq 2$~GeV, in favour of a larger 
value of $\alpha (M_Z^2)^{-1} = 128.978 \pm 0.027$.  
As a consequence the predicted value of the mass of the 
(Standard Model) Higgs boson is increased, so that the preferred value is close to the LEP2 
bound.
\end{abstract}

\vspace*{0.5cm}
The value of the QED coupling at the $Z$ pole, $\alpha(M_Z^2)$, is the poorest known of 
the three parameters ($G_F, M_Z, \alpha (M_Z^2)$) which define the standard electroweak 
model.  Indeed it is the precision to which we know $\alpha (M_Z^2)$ which limits the 
accuracy of the indirect prediction of the mass $M_H$ of the (Standard Model) Higgs boson 
\cite{Z,GURTU}.  In fact the predicted allowed domain of $M_H$ 
appears to be not too far from being in conflict with the 
direct LEP2 bound on the mass.
However the quoted mass range for the Higgs does not allow for uncertainty
in the value of $\alpha (M_Z^2)$ itself, and so the conflict is
less severe than implied by the $\chi^2$ profiles versus $M_H$ \cite{Z,GURTU}.
Clearly a more precise determination of $\alpha (M_Z^2)$ is especially 
important.

The value of $\alpha (M_Z^2)$ is obtained from
\be
\label{eq:a1}
\alpha^{-1} \; \equiv \; \alpha (0)^{-1} \; = \; 137.03599976 (50)
\ee
using the relation
\be
\label{eq:a2}
\alpha (s)^{-1} \; = \; \left ( 1 - \Delta \alpha_{\rm lep} (s) - \Delta \alpha_{\rm had}^{(5)} (s) 
- \Delta \alpha^{\rm top} (s) \right ) \: \alpha^{-1},
\ee
where the leptonic contribution to the running of the $\alpha$ is known to 3 loops \cite{ST}
\be
\label{eq:a3}
\Delta \alpha_{\rm lep} (M_Z^2) \; = \; 314.98 \times 10^{-4}.
\ee
From now on we omit the superscript (\ref{eq:a5}) on $\Delta \alpha_{\rm had}$ and assume that it 
corresponds to five flavours.  We will include the contribution of the sixth flavour, 
$\Delta \alpha^{\rm top} (M_Z^2) = -0.76 \times 10^{-4}$, at the end.  To determine the 
hadronic contribution we need to evaluate
\be
\label{eq:a4}
\Delta \alpha_{\rm had} (s) \; = \; - \frac{\alpha s}{3 \pi} \: P \int_{4m_\pi^2}^{\infty} \: 
\frac{R (s^\prime) ds^\prime}{s^\prime (s^\prime - s)}
\ee
at $s = M_Z^2$, where $R = \sigma (e^+ e^- \rightarrow {\rm hadrons})/\sigma (e^+ e^- 
\rightarrow \mu^+ \mu^-)$.

The early determinations of $\Delta \alpha_{\rm had} (M_Z^2)$ made maximum use of the 
$e^+ e^-$ measurements of $R(s)$, using the sum of the exclusive hadronic channels $(e^+ 
e^- \rightarrow 2\pi, 3\pi, \ldots K\bar{K}, \ldots)$ for $\sqrt{s} \lapproxeq 1.5$~GeV and the 
inclusive measurement of $\sigma (e^+ e^- \rightarrow {\rm hadrons})$ at larger energies, 
culminating in the analysis of Eidelman and Jegerlehner \cite{EJ} which gave
\be
\label{eq:a5}
\Delta \alpha_{\rm had} (M_Z^2) \; = \; (280.4 \pm 6.4) \: \times \: 10^{-4}.
\ee
Other similar determinations, which gave compatible results, can be found in 
Refs.~\cite{BP,S,ADH}, where the latter analysis incorporated information from $\tau 
\rightarrow$ (hadrons $+ \nu_\tau$) decays.  However, in the last few years the 
determinations of $\Delta \alpha_{\rm had}$ from (\ref{eq:a4}) have relied more and more 
on theoretical input.  First, perturbative QCD was used to better describe $R(s)$ (for $\sqrt{s} 
> 3$~GeV) in continuum energy regions above the resonances up to the next flavour threshold \cite{MZ}.  
Then, encouraged by the success of perturbative QCD to describe $\tau$ decay, it was used 
down to 1.8~GeV, across a region with sparse data on $R(s)$, giving \cite{KS}
\be
\label{eq:a6}
\Delta \alpha_{\rm had} (M_Z^2) \; = \; (277.5 \pm 1.7) \times 10^{-4},
\ee
where $\pm 10^{-4}$ comes from the uncertainty in perturbative QCD.  A recent analysis \cite{MOR}, 
which follows the more conservative perturbative QCD input of \cite{MZ}, finds
\bea
\label{eq:a7}
\Delta \alpha_{\rm had} (M_Z^2) & = & (274.18 \pm 2.52) \: \times \: 10^{-4} \quad ({\rm 
inclusive}) \\
\label{eq:a8}
\Delta \alpha_{\rm had} (M_Z^2) & = & (276.97 \pm 2.90) \: \times \: 10^{-4} \quad ({\rm 
exclusive})
\eea
according to whether $R$ is evaluated in the interval $1.46 < \sqrt{s} < 2.125$~GeV using 
either the inclusive data for $e^+ e^- \rightarrow$ hadrons or the sum of the data for the 
exclusive channels.  The difference in the input in this region can be seen in Fig.~1.

\begin{figure}[htb]
\label{fig:fig1}
\begin{center}
\mbox{\epsfig{figure=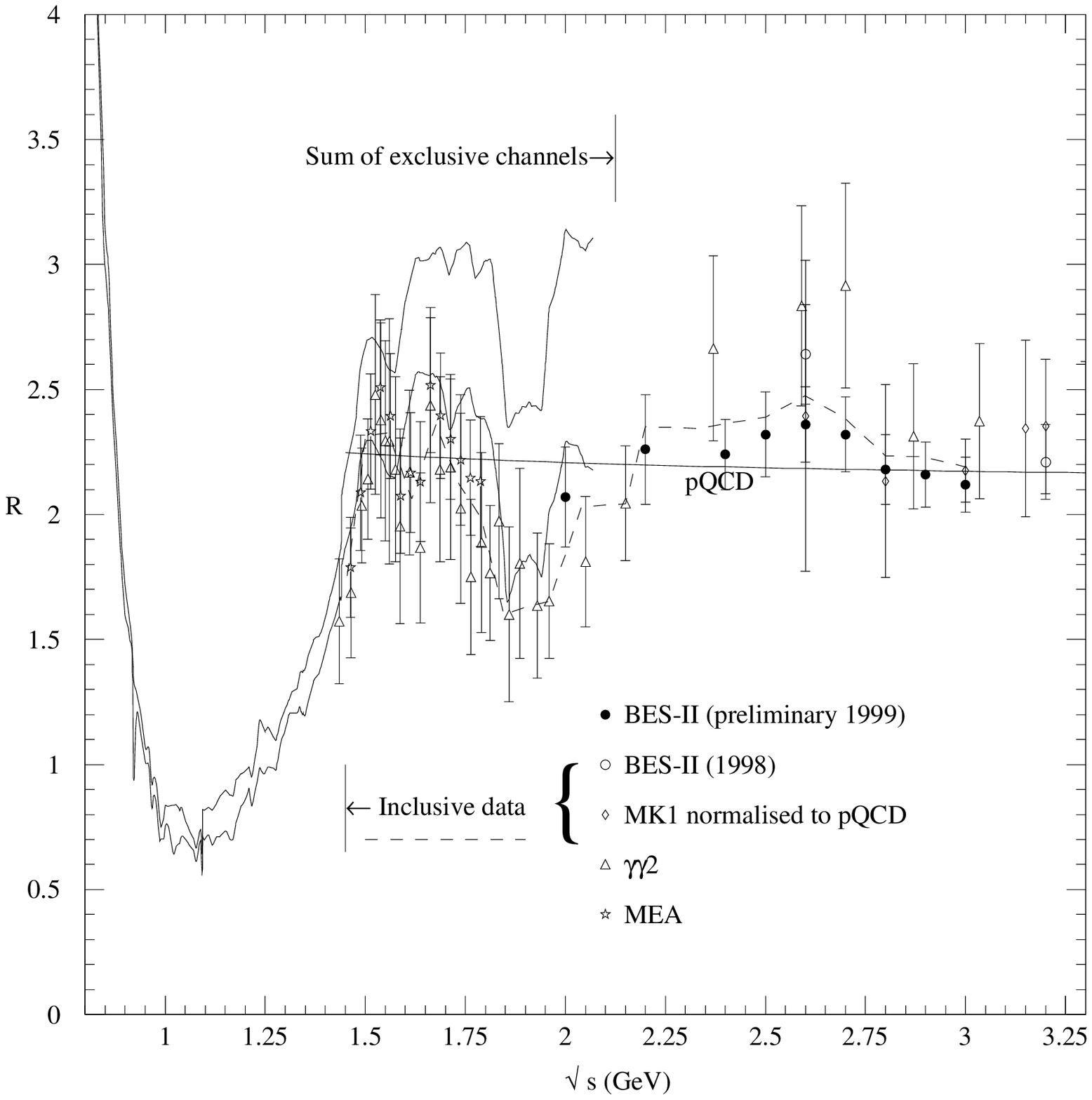,width=17.cm}}
\caption{A plot of the ratio $R(s)$ versus $\sqrt{s}$ in the most sensitive $\sqrt{s}$ interval.  
Up to $\sqrt{s} = 2.125$~GeV we show by continuous lines the upper and lower bounds of 
the sum of the exclusive channels.  However in the \lq\lq exclusive\rq\rq\ analysis we only use these data 
up to $\sqrt{s} = 1.9$~GeV.  The value quoted in Table~1 corresponds to the average of the 
bounds.  Above $\sqrt{s} = 1.46$~GeV we show the inclusive measurements of $R(s)$, 
together with the interpolation (dashed curve) used in (\ref{eq:a4}).  For $2.8 < \sqrt{s} < 
3.74$ perturbative QCD is used to evaluate $R$; the central QCD prediction is shown by the continuous 
curve.  In the region $1.46 < \sqrt{s} < 1.9$ we used, in turn, the inclusive and exclusive data to evaluate 
$\Delta \alpha_{\rm had}$, see Table~1.
}
\end{center}
\end{figure}

Other recent calculations of $\alpha (M_Z^2)$, based on (\ref{eq:a4}), have used the analytic 
behaviour in the complex $s$ plane in attempts to reduce the error coming from the 
measurements of $R$, see, for example, Refs.~\cite{GRO}--\cite{DH}.  These methods have 
been critically reviewed by Jegerlehner \cite{J}.  In this work he showed that a space-like 
evaluation of (\ref{eq:a4}) at $s = - (2.5~{\rm GeV})^2$, followed by an analytic 
continuation \cite{EJKV} to $s = - M_Z^2$, and then around the semicircle to $s = M_Z^2$, 
gives
\be
\label{eq:a9}
\Delta \alpha_{\rm had} (M_Z^2) \; = \; (277.82 \pm 2.54) \: \times \: 10^{-4},
\ee
in agreement with the direct determination of (\ref{eq:a4}) at $s = M_Z^2$.  He concludes that no appreciable 
reduction of the error, arising from the data, is obtained by the analytic methods.  The 
reduction in the error in (\ref{eq:a7}--\ref{eq:a9}), in comparison to that in (\ref{eq:a5}), is 
due to the use of perturbative QCD to evaluate $R$ in energy regions where it is expected to 
be reliable.

We see that the values (\ref{eq:a6}), (\ref{eq:a8}) and (\ref{eq:a9}) are in good agreement with each other, 
but that the determination (\ref{eq:a7}) using inclusive data below 2.1~GeV is significantly 
lower.  This discrepancy has existed for some time, although most recent determinations have 
relied on the sum of the exclusive data.  However BES-II 1999 data have recently been 
presented \cite{ZHAO} which, although preliminary, already clarify the behaviour of 
$R(s)$ in the critical $2 \lapproxeq \sqrt{s} \lapproxeq 3$~GeV domain.  These data update 
the previous, published, BES-II measurements \cite{BES}.  They are shown by the solid circular data points in 
Fig.~1.  First, we see that in the region $\sqrt{s} \sim 2.5$~GeV the new measurements are 
closer to the expectations of perturbative QCD than the previous measurements.  Second, at 
$\sqrt{s} = 2-2.2$~GeV they join reasonably smoothly with the inclusive measurements of 
the $\gamma\gamma 2$ collaboration \cite{GG2}, whereas they lie in the region of, or below, the lower 
limit of the determination of $R$ from the sum of the exclusive channels.

Here we re-evaluate (\ref{eq:a4}) at $s = M_Z^2$, incorporating these new BES-II
data.
To be 
specific we use the dashed curve for $\sqrt{s} > 1.46$~GeV for the so-called \lq\lq inclusive\rq\rq\ 
determination of $\Delta \alpha_{\rm had} (M_Z^2)$, and the average of the bounds for $\sqrt{s} 
< 1.9$~GeV for the \lq\lq exclusive\rq\rq\ determination.  In addition we use the data, and follow the 
method, described in \cite{MZ,MOR}.  The errors on the input \lq data\rq\ values of $R(s^\prime)$ 
are calculated using a correlated $\chi^2$ minimization to combine different data sets, as described 
in detail in Ref.~\cite{ADH}.

The data from $\sqrt{s} = 2.8$~GeV up to the ${\rm D\overline{D}}$ charm meson threshold are 
in good agreement\footnote{The MARK I data \cite{MI} are normalized as described in \cite{MZ}.} with 
perturbative QCD, apart, of course, in $J/\psi$ and $\psi^\prime$ resonance regions.  In the 
energy intervals where perturbative QCD is believed to be valid 
($2.8 < \sqrt{s} < 3.74$~GeV and $\sqrt{s} > 5$~GeV) we use both the two-loop expression 
with the quark mass explicity 
included and the massless three-loop expression \cite{CHET} calculated in the 
$\overline{\rm MS}$ renormalization scheme\footnote{The uncertainty due to using a 
different scheme may be estimated to be of the order of the ${\cal O} (\alpha_S^4)$ 
correction, which is about $3 \sum e_q^2 r_3 (\alpha_S/\pi)^4$.  We may take $r_3 = -128$ 
\cite{KS2} which leads to an uncertainty much smaller than that given in Table 1.}.  In 
addition to the above, we include the perturbative QCD error\footnote{We thank Thomas Teubner 
for valuable discussions concerning the perturbative QCD contribution.} coming from varying $m_c, 
m_b, M_Z$ within the uncertainties quoted in the \cite{PDG}, $\alpha_S (M_Z^2) = 0.119 
\pm 0.002$ and varying the scale $\alpha_S (cs)$ in the range $0.25 < c < 4$.  The contributions 
to $\Delta \alpha_{\rm had} (M_Z^2)$ of (\ref{eq:a4}) from the various $s^\prime$ intervals are 
listed in Table~1.

The recent Novosibirsk measurements have considerably improved the exclusive data below 
$1.4$~GeV, see, for example, Refs.~\cite{RHO,NOV}.  For instance, the measurements of $e^+ e^- 
\rightarrow \pi^+ \pi^-$ \cite{RHO} are of much improved precision such that, when taken 
together with chiral perturbation theory estimates, we find that the error on the $2\pi$ 
contribution to $\Delta \alpha_{\rm had}$ for $\sqrt{s} < 0.96$~GeV is reduced to about $\pm 0.5 
\times 10^{-4}$.  We checked that our results for the 
contributions from the exclusive channels were in agreement with the detailed table of results 
given in Ref.~\cite{ADH}, if we were to omit the new Novosibirsk data.

From Table~1 we see that the largest uncertainties in $\Delta \alpha_{\rm had}$ occur in the 
$1.46 < \sqrt{s^\prime} < 2.8$~GeV interval, although the introduction of the new BES-II data 
\cite{ZHAO} have led to a significant improvement.  It is interesting to note that if we were 
to assume that perturbative QCD gave a reliable prediction for $R(s^\prime)$ in the region 
$1.9 < \sqrt{s^\prime} < 2.8$~GeV, then the contribution from this interval would be $(13.18 \pm 
0.09) \times 10^{-4}$.  This is essentially the same as the \lq\lq inclusive\rq\rq\ contribution, 
but with a much smaller error.  The \lq\lq exclusive\rq\rq\ contribution is a little larger on account 
of the necessity to smoothly match up to the \lq\lq exclusive\rq\rq\ data at $\sqrt{s} = 1.9$~GeV.

\begin{table}[htb] 
\caption{Contributions to $\Delta \alpha_{\rm had} (M_Z^2) \times 10^4$ of (\ref{eq:a4})  
coming from the different $\sqrt{s^\prime}$ intervals. The alternative values in the round 
brackets use the summation of exclusive channels as the contribution from the region 
1.46--1.9~GeV, rather than that from the inclusive data for $R(s^{\prime})$.  Perturbative 
QCD is used to evaluate the contributions in the intervals $2.8 < \sqrt{s^\prime} < 3.74$ and 
$\sqrt{s^\prime} > 5$~GeV.  The errors on these contributions are described in the text.}
\begin{center} 
\begin{tabular}{|c|c|} \hline 
$\sqrt{s^\prime}$ interval & Contribution to  \\ 
(GeV) & $\Delta \alpha_{\rm had} \times 10^4$ \\ \hline 
$2m_{\pi} - 1.46^{a}$ & 38.41 $\pm$ $\left\{\begin{array}{c}0.52 \\ 0.60^{b} 
\end{array}\right\}$  \\ 
1.46 - 1.9 & 8.66 $\pm$ 0.60$^{c}$ \\ 
& $\left(10.32 \pm 1.06^{b}\right) $ \\ 
1.9 - 2.8 & 13.24 $\pm$ 0.88$^{c}$ \\ 
& $\left(13.88 \pm 0.88\right)$ \\
2.8 - 3.74 & 9.73 $\pm$ 0.05$^d$ \\ 
3.74 - 5 &  15.02 $\pm$ 0.49  \\  
5 - $\infty$ & 169.97 $\pm$ 0.64$^d$  \\ 
$\omega$, $\phi$, $\psi$'s, $\Upsilon$'s & 18.79 $\pm$ 0.58 \\ \hline 
$\Delta \alpha_{\rm had}^{(5)} \times 10^4$ &  273.82 $\pm$ 1.97  \\  
& $\left( 276.12 \pm  2.20 \right) $ \\ \hline  
$\alpha^{-1} (M_Z^2)$ & $128.978 \pm 0.027$ \\
& $(128.946 \pm 0.030)$ \\ \hline
\end{tabular} 
\end{center}
\scriptsize{$^{a}$ The upper (lower) error corresponds to the $2\pi$ (remaining) exclusive 
channels.} 

\scriptsize{$^{b,c,d}$ Errors with identical superscripts are added linearly.  The remaining 
errors are added in quadrature.}

\scriptsize{$^d$ For the pQCD contribution we take the mass of the charm quark $m_c = 1.46$~GeV, 
and the scale $\mu^2$ of the QCD coupling $\alpha_S$ to be $s^\prime$.}
\end{table}

To conclude, we see that the introduction of new data has improved the error of $\Delta \alpha_{\rm had}$, 
and moved the \lq\lq inclusive\rq\rq\ and \lq\lq exclusive\rq\rq\ determinations a little closer together 
(compare (\ref{eq:a7}) and (\ref{eq:a8}) with the values of $\Delta \alpha_{\rm had}$ in Table~1).  
Second, the new BES-II data appear to join more smoothly to the inclusive data than the exclusive 
measurements.  However the selection is not conclusive and there remains a major residual uncertainty 
in the $1.5 < \sqrt{s} < 1.9$~GeV interval, which is reflected by the two different results for 
$\Delta \alpha_{\rm had}$ listed in Table~1.  However if we take the favoured \lq\lq inclusive\rq\rq\ result
\be
\label{eq:a10}
\Delta \alpha_{\rm had}^{(5)} \; = \; (273.82 \pm 1.97) \: \times \: 10^{-4},
\ee
then
\be
\label{eq:a11}
\alpha^{-1} (M_Z^2) \; = \; 128.978 \pm 0.027.
\ee
Using this value, the latest $\chi^2$ profile, obtained\footnote{We thank Martin Gr\"unewald for 
making this plot.} by the LEP and SLD Electroweak Working Group \cite{GURTU,PIETRZYK}, 
for different values of the 
mass of the (Standard Model) Higgs boson, is shown by the dashed line in Fig.~2.  We see that it produces a 
significant increase in the predicted mass of the Higgs, with the preferred value lying close to 
the LEP2 bound.  The new $\chi^2$ profile accommodates 
the LEP2 bound on the mass more comfortably.  Note that even the determination using the exclusive data 
up to $\sqrt{s} = 1.9$~GeV, now predicts a {\it lower} value of $\Delta \alpha_{\rm had}$, 
\be
\label{eq:a12}
\Delta \alpha_{\rm had}^{(5)} \; = \; (276.12 \pm 2.20) \: \times \: 10^{-4},
\ee
and a {\it higher} Higgs mass than before, with the corresponding $\chi^2$ profile curve being nearer 
to the dashed curve than the original continuous curve in Fig.~2.

We emphasize that these standard $\chi^2$ profiles versus $M_H$ do not include
the uncertainty due to $\Delta\alpha_{\rm had}$. Clearly when confidence limits
are placed on the Higgs mass, it is important that the uncertainty due to
$\Delta\alpha_{\rm had}$ is included.

After the completion of this work we were made aware of a recent very preliminary 
determination \cite{PIETRZYK},
\be
\label{eq:a13}
\Delta\alpha_{\rm had}^{(5)} \; = \; (275.5\pm4.6) \: \times \: 10^{-4},
\ee
which also incorporates the BES-II data. The details of the calculation are
not available, but from the choice of input for $R$ in the region
$\sqrt{s}\lapproxeq2$~GeV we would anticipate a value of $\Delta\alpha_{\rm had}$
somewhat closer to our ``exclusive'' determination, given in (\ref{eq:a12}).
Therefore, if we assume the contributions of the other regions are the same,
the two results (\ref{eq:a12}) and (\ref{eq:a13}) are in good agreement
with each other.
\begin{figure}[htb]
\label{fig:fig2}
\begin{center}
\mbox{\epsfig{figure=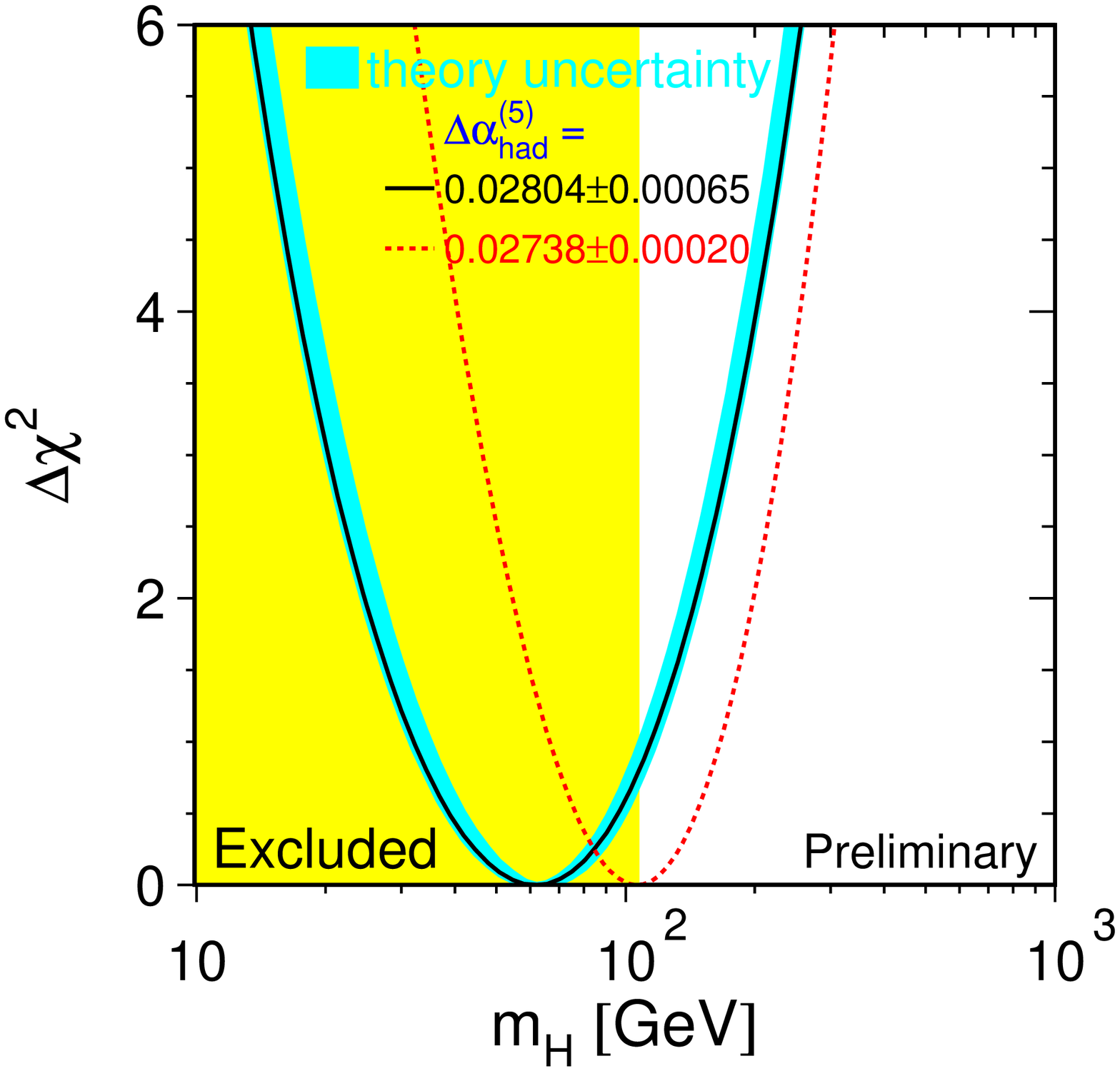,width=17.cm}}
\caption{
The latest $\chi^2$ profile versus the Standard Model Higgs mass 
obtained by the LEP and SLD Electroweak 
Working Group \cite{GURTU,PIETRZYK}, which includes the latest preliminary experimental data, 
using $\Delta \alpha_{\rm had}$ of (\ref{eq:a5}) compared to that obtained 
using (\ref{eq:a10}), shown by continuous and dashed curves respectively.
Note that the $\chi^2$ profiles do not include the uncertainty of $\Delta\alpha_{\rm had}$.
The shaded region to the left is excluded by searches for the Higgs boson at LEP2.
}
\end{center}
\end{figure}

\section*{Acknowledgements}

We thank Martin Gr\"unewald,
Andrei Kataev, Thomas Teubner and Zhengguo Zhao for informative discussions.  
One of us (MGR) thanks the Royal Society for 
support. \\

\end{document}